\documentclass[aps,prb,preprint,bibliography]{revtex4-1}
\usepackage[utf8]{inputenc}
\usepackage{amsmath}

\usepackage{mathtools}
\usepackage{amsfonts}
\usepackage{amssymb}
\usepackage{graphicx}
\usepackage{textcomp}
\usepackage{color}
\usepackage[dvipsnames]{xcolor}
\usepackage{comment}

\newcommand{\rcaps}[1]
    {\MakeUppercase{\romannumeral #1}}

\begin{document}

\title{Non-Linear Hall Effect in Multi-Weyl Semimetals}

\author{Saswata Roy}
\affiliation{Undergraduate Programme, Indian Institute of Science, Bangalore 560012, India}
\author{Awadhesh Narayan}
\email{awadhesh@iisc.ac.in}
\affiliation{Solid State and Structural Chemistry Unit, Indian Institute of Science, Bangalore 560012, India}

\date{\today}

\begin{abstract}
In the presence of time reversal symmetry, a non-linear Hall effect can occur in systems without an inversion symmetry. One of the prominent candidates for detection of such Hall signals are Weyl semimetals. In this article, we investigate the Berry curvature induced second and third order Hall effect in multi-Weyl semimetals with topological charges $n =1, 2, 3$. We use low energy effective models to obtain general analytical expressions and discover the presence of a large Berry curvature dipole in multi-Weyl semimetals. We also study the Berry curvature dipole in a realistic tight-binding lattice model and observe two different kinds of variation with increasing topological charge -- these can be attributed to different underlying Berry curvature components. We provide estimates of the signatures of second harmonic of Hall signal in multi-Weyl semimetals, which can be detected experimentally. Furthermore, we predict the existence of a third order Hall signal in multi-Weyl semimetals. We derive the analytical expressions of Berry connection polarizability tensor, which is responsible for third order effects, using a low energy model and estimate the measurable conductivity. Our work can help guide experimental discovery of Berry curvature multipole physics in multi-Weyl semimetals.
\end{abstract}

\maketitle

\section{Introduction}

A geometric phase may be acquired by a classical or quantum system when undergoing cyclic adiabatic processes. Independently discovered by Kato~\cite{kato1950adiabatic}, Pancharatnam~\cite{pancharatnam1956generalized}, and Longuet-Higgins \textit{et al.}~\cite{longuet1958studies}, in different settings, the notion was organized in a general framework by Berry~\cite{berry1984quantal}. Berry's phase plays a key role in various condensed matter phenomena such as quantum Hall effect, spin hall effects, electric polarization and orbital magnetism~\cite{RevModPhys.82.1959}. Berry curvature, which originates from the Berry flux, acts as the magnetic field in the momentum space and is intimately connected to the celebrated quantum Hall effect~\cite{PhysRevLett.49.405}.

Recently, Sodemann and Fu, building on earlier work~\cite{deyo2009semiclassical,moore2010confinement}, have discovered the role of the \emph{first order moment} of the Berry curvature -- termed Berry curvature dipole (BCD) -- in transport properties of quantum systems~\cite{PhysRevLett.115.216806}. Surprisingly, BCD can lead to a non-linear Hall effect in time-reversal invariant systems~\cite{PhysRevLett.115.216806}. The conventional Hall effect can be thought of as connected to the zeroth order moment of the Berry curvature while the first order moment leads to the second harmonic generation in the Hall signal~\cite{du2021perspective,ortix2021nonlinear}. A growing number of generalizations to other non-linear phenomena have also been proposed~\cite{hamamoto2017nonlinear,araki2018strain,konig2019gyrotropic,papaj2019magnus,yu2019topological,zeng2019nonlinear,mandal2020magnus,nakai2019nonreciprocal,zeng2020fundamental,PhysRevB.104.115420,PhysRevB.100.195117,du2019disorder,du2021quantum,xiao2019theory,resta2021linear}. Similarly, higher order moments of the Berry curvature are predicted to produce higher harmonics of the Hall signal~\cite{zhang2020higher}. To completely describe the third order conductivity tensor, another intrinsic band geometric quantity known as Berry connection polarizability (BCP) tensor has been proposed~\cite{liu2021berry}. 

BCDs and the related non-linear Hall effect have been predicted for a spectrum of materials ranging from two-dimensional materials~\cite{low2015topological,PhysRevB.98.121109,zhang2018electrically,joseph2021topological,samal2021nonlinear}, tilted massive Dirac and Weyl cone systems~\cite{PhysRevLett.121.266601,xiao2020electrical,singh2020engineering}, Weyl semimetals~\cite{zhang2018berry}, large Rashba systems~\cite{facio2018strongly}, and strained monolayer and bilayer graphene~\cite{PhysRevLett.123.196403}, to name just a few. In parallel to theoretical predictions, there have been noteworthy developments along the experimental front. Second order non-linear Hall effect was experimentally first observed in few layer transition metal dichalcogenide WTe$_2$~\cite{Ma2019,kang2019nonlinear}. Following these pioneering reports, signatures of non-linear Hall signals have been detected in a growing number of materials, including Dirac semimetals~\cite{shvetsov2019nonlinear}, Kondo materials~\cite{dzsaber2021giant}, artificially corrugated bilayer graphene~\cite{Ho2021}, monolayer WSe$_2$~\cite{qin2021strain}, twisted WSe$_2$~\cite{huang2020giant}, MoTe$_2$~\cite{tiwari2021giant}, organic Dirac materials~\cite{kiswandhi2021observation}, topological insulator surfaces~\cite{he2021quantum} and Weyl semimetal TaIrTe$_4$~\cite{kumar2021room}. Very recently, third order non-linear signals have been experimentally detected in MoTe$_2$~\cite{Lai2021}.    

An important class of materials for the realization of non-linear Hall effect are Weyl semimetals -- topological semimetallic systems exhibiting a Weyl fermion dispersion in the low energy regime~\cite{rao2016weyl,yan2017topological,armitage2018weyl}. The point where two bands cross is known as the Weyl point. These points act as a source or sink of Berry curvature in momentum space and hence are associated with integer charged monopole and chirality. Weyl points with opposite chirality can merge and annihilate each other. On the other hand, Weyl nodes with same chirality can merge to form Weyl points with higher topological charge, which are named multi-Weyl semimetals. These systems are stabilised only when point group symmetry protects such a merging~\cite{PhysRevLett.108.266802}, hence there are discrete allowed values of topological charge, $n$~\cite{huang2016new,xu2011chern}. Multi-Weyl semimetals have been predicted to show many interesting properties such as chiral effects, anomalous transport phenomena and distinct optical signatures~\cite{ahn2017optical,ahn2016collective,sinha2019transport,kulikov2020josephson,park2017semiclassical,huang2017topological,dantas2020non,lu2019quantum,gorbar2017anomalous,mukherjee2018doping,sun2017rkky,soto2020dislocation,chowdhury2021light,wang2017quantum,nag2020thermoelectric}.

Motivated by these exciting developments, here we study BCD in multi-Weyl semimetals using a low energy model, as well as a suitable tight-binding model. We obtain general analytical expressions (for arbitrary $n$) for the Berry curvature using the low energy model and discover the presence of a large BCD in multi-Weyl semimetals. To complement our analytical calculations, we next turn to a three dimensional tight-binding model and study the dependence of BCD on different model parameters for different monopole charges. In general, we find that a higher monopole charge facilitates higher magnitude of generated second harmonic of Hall signal. Further, we calculate the BCP tensor components for our low energy model and predict the existence of third-order Hall conductivity in multi-Weyl semimetals. For the existing symmetries in our system the third order contribution is less than that of second order contribution, but can be dominant if second order signal is suppressed. Our work can guide experimental discovery of Berry curvature multipole physics in multi-Weyl semimetals and also help to characterize new classes of multi-Weyl semimetal materials with higher topological charges.

\section{Berry curvature, Berry curvature dipole and Berry connection polarizability tensor}

Here we briefly review the notions of Berry curvature, its dipole (BCD) and BCP tensor, and their relation to transport properties. In response to an oscillating electric field, $\mathbf{E}(t) = \mathrm{Re}[\boldsymbol{\varepsilon} e^{i\omega t}]$, a non-linear current, $J_a = \mathrm{Re}[J_a ^{(0)} + J_a ^{(2)} e^{2i\omega t}]$ flows through the material with $J_a^{(0)} = \chi_{abc}^{(0)}\varepsilon_b\varepsilon_c^*$ and $J_a^{(2)} = \chi_{abc}^{(2)}\varepsilon_b\varepsilon_c$. In systems that preserve time-reversal symmetry the coefficients are given by~\cite{PhysRevLett.115.216806}

\begin{equation}
    \chi_{abc}^{(0)} = \chi_{abc}^{(2)} = \frac{\epsilon^{acd} D_{bd} e^3 \tau}{2 \hslash^2 (1+i\omega t)},
    \label{eqn:2nd_order_conductivity}
\end{equation}

where $\epsilon^{acd}$ is the Levi-Civita symbol, $\tau$ is momentum relaxation time, $\omega$ is the frequency and $-e$ is the charge of electron. Here $D_{bd}$ is the BCD, which is formulated as

\begin{equation}
    D_{bd} = -\sum_{i}\int [d\boldsymbol{k}] \frac{\partial \epsilon_{\boldsymbol{k}}^i}{\partial k_b} \boldsymbol{\Omega}_{i\boldsymbol{k}}^d \frac{\partial f_{\boldsymbol{k}}}{\partial \epsilon_{\boldsymbol{k}}^i}.
    \label{eqn:berry_dipole}
\end{equation}

Here $a,b,c,d \in \{x,y,z\}$, $f_{\boldsymbol{k}}$ is the equilibrium Fermi-Dirac distribution, $\epsilon_{\boldsymbol{k}}^i$ is the energy of the $i$-th energy band and wavevector $\boldsymbol{k}$, $\Omega_{i\boldsymbol{k}}^d$ is the Berry curvature component in direction $d$ for the $i$-th energy band and wave vector $\boldsymbol{k}$ and $[d\boldsymbol{k}] = d^3\boldsymbol{k}/{(2\pi)^3}$.  The Berry curvature, in turn, can be found as~\cite{RevModPhys.82.1959}

\begin{equation}
	    \Omega_{i\boldsymbol{k}}^a =-\epsilon^{abc} \text{Im} \sum_{j\neq i}\frac{\langle i|\frac{\partial H}{\partial k_b}|j\rangle \langle j|\frac{\partial H}{\partial k_c}|i\rangle-\langle i|\frac{\partial H}{\partial k_c}|j\rangle \langle j|\frac{\partial H}{\partial k_b}|i\rangle} {(\epsilon_{\boldsymbol{k}}^i-\epsilon_{\boldsymbol{k}}^j)^2},
	    \label{eqn:berry_curvature}
	\end{equation}
	
where $|i\rangle$ is an eigenstate of the Hamiltonian $H$ that corresponds to band $i$ with energy $\epsilon_{\boldsymbol{k}}^i$ for a wave vector $\boldsymbol{k}$. Notably, the integral in Equation \ref{eqn:berry_dipole} can survive in time-reversal symmetric systems, as long as the inversion symmetry is broken -- this enables a finite Berry curvature to obtain a non-zero value of the BCD. We also define a BCD \emph{density}~\cite{PhysRevLett.121.266601}

\begin{equation}
    d_{bd}(\boldsymbol{k}) = (\partial_{\boldsymbol{k}}^b \epsilon_{\boldsymbol{k}})\Omega_{\boldsymbol{k}}^d,
    \label{eqn:dipole_density}
\end{equation}

which is the kernel of the integral in Equation \ref{eqn:berry_dipole}. As we will see, this quantity, $d_{bd}$ gives further insights into the physics of the Berry curvature dipole.

Going beyond the second order, third order contributions to the Hall effect can be connected to the BCP tensor, which is an intrinsic band geometric quantity~\cite{liu:2021}. It can be expressed in terms of the unperturbed eigenstates $ |i\rangle$ and band energy $\epsilon_i$ as~\cite{liu:2021}

\begin{equation}
    G_{ab} = 2 \mathrm{Re} \sum_{i \neq j} \frac{(\mathcal{A}_a)_{ij}(\mathcal{A}_b)_{ij}}{\epsilon_i - \epsilon_j}.
    \label{eqn:BPT}
\end{equation}

where, $\boldsymbol{\mathcal{A}}$ is the Berry connection. The Berry curvature needs to be corrected to first order in presence of an external electric field

\begin{equation}
    \tilde{\boldsymbol{\Omega}}(\boldsymbol{k}) = \boldsymbol{\Omega}(\boldsymbol{k}) +\boldsymbol{\Omega}^{(1)}(\boldsymbol{k}),
    \label{eqn:omega_bpt}
\end{equation}

which is given by

\begin{equation}
    \boldsymbol{\Omega}^{(1)} = \nabla_{\boldsymbol{k}}\times \boldsymbol{\mathcal{A}}^{(1)}.
    \label{eqn:omega_corr_1}
\end{equation}

In turn, the first order correction of Berry connection, $\boldsymbol{\mathcal{A}}^{(1)}$, is related to the BCP tensor as

\begin{equation}
    \mathcal{A}_a^{(1)} (\boldsymbol{k}) = G_{ab} (\boldsymbol{k}) E_b.
    \label{eqn:flux correction_1}
\end{equation}

Using Equations \ref{eqn:omega_bpt} and \ref{eqn:flux correction_1}, we can also define a Berry curvature polarizability as

\begin{equation}
    P_{ab} = \frac{\partial \tilde{\Omega}_a}{\partial E_b} = \epsilon_{acd} \partial_c G_{db}.
    \label{eqn:bc_polarizability}
\end{equation}

Using the components of the BCP tensor, we can calculate the third order conductivity tensor $\chi$ that follows $j_a^{(3)} = \chi_{abcd} E_b E_c E_d$, with Einstein summation convention assumed. Further, $\chi$ can be broken into two parts -- $\chi^{\rcaps{1}}$ which is linear in $\tau$ and $\chi^{\rcaps{2}}$ which is proportional to $\tau^3$.

\begin{subequations}
    \begin{equation}
        \chi^{\rcaps{1}}_{abcd} = \tau\left[2\int [d\boldsymbol{k}](\partial_a \partial_b G_{cd}) f_{\boldsymbol{k}} - \int [d\boldsymbol{k}](\partial_c \partial_d G_{ab}) f_{\boldsymbol{k}} - \frac{1}{2} \int [d\boldsymbol{k}](v_a v_b G_{cd}) f_{\boldsymbol{k}}^{\prime\prime}\right],
    \end{equation}
    \begin{equation}
        \chi^{\rcaps{2}}_{abcd} = - \tau^3 \int [d\boldsymbol{k}]v_a \partial_b \partial_c \partial_d f_{\boldsymbol{k}}.
    \end{equation}
\end{subequations}

Here $v_i = \frac{\partial E}{\partial{k_i}}$ and $\chi^{\rcaps{1}}_{abcd}$ is the major contributor to the third order non-linear Hall response as it has a linear dependence on the relaxation time $\tau$.

\section{Results and Discussion}

\subsection{Low-energy model}

\begin{figure}
    \centering
	 \includegraphics[width=0.94\linewidth]{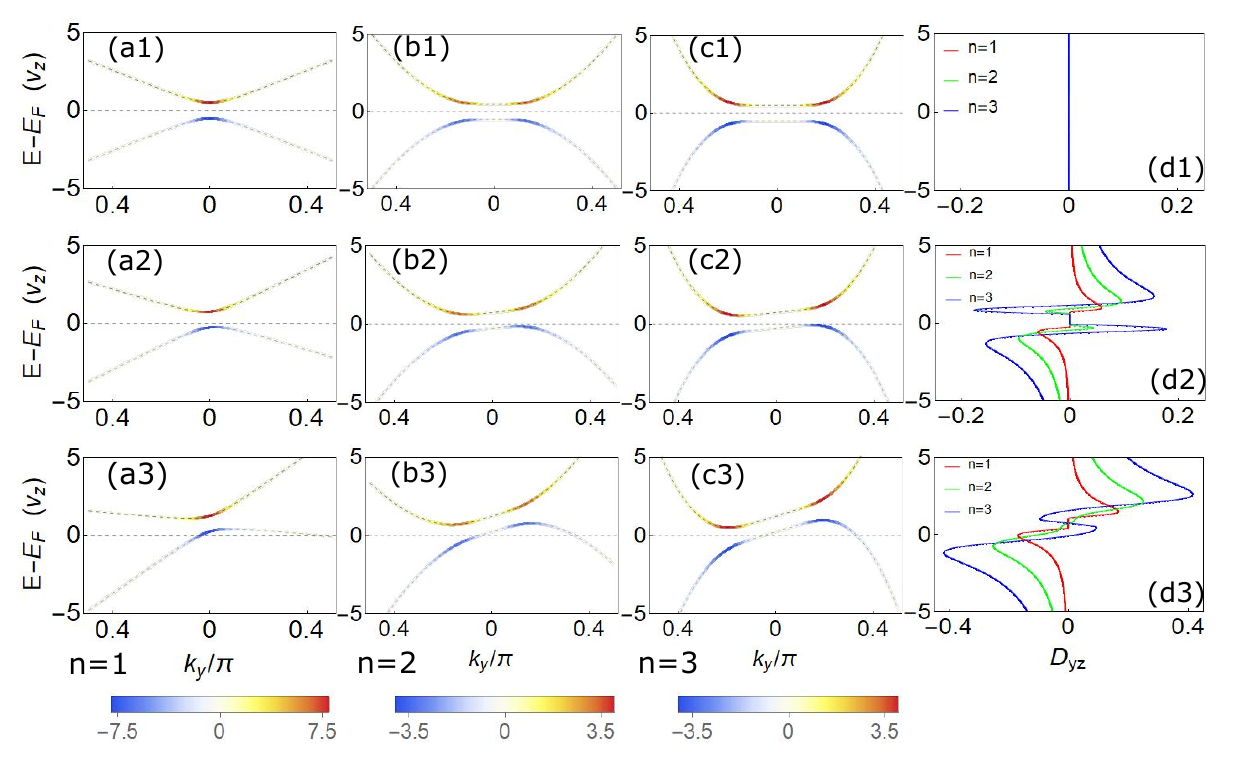}
	  \caption{\textbf{Band structure, Berry curvature and Berry curvature dipole in multi-Weyl semimetals.} Band structures of the low energy Hamiltonian with tilt for multi-Weyl semimetals along $k_y$ (setting $k_x=k_z=0$) with the following choice of parameters (a1) $n=1$, $C_s=0$, (a2) $n=1$, $C_s=0.5$, (a3) $n=1$, $C_s=1.5$, (b1) $n=2$, $C_s=0$, (b2) $n=2$, $C_s=0.5$, (b3) $n=2$, $C_s=1.5$, (c1) $n=3$, $C_s=0$, (c2) $n=3$, $C_s=0.5$, (c3) $n=3$, $C_s=1.5$. Value of the Berry curvature is superimposed on the band structure and corresponding color scales are shown. The Berry curvature is concentrated close to the multi-Weyl nodes, with opposite values for conduction and valence bands. (d1-d3) Berry curvature dipole for the corresponding tilt ($C_s$) values with $n=1$ in red, $n=2$ in green and $n=3$ in blue. Berry curvature dipole becomes non-zero only when a tilt is introduced, as expected. Note the increase in dipole with increasing topological charge. The maximum value of dipole is observed in the vicinity of the energy where Berry curvature is most concentrated and this maximum value increases with an increase in $C_s$. Other parameters common to all plots are $v_z=1$, $\alpha=2$, $s=-1$, and $Q=0.5$.}
	   \label{fig:band_bcd}
\end{figure}

We begin with a low energy model Hamiltonian~\cite{Dantas2018, menon:2019} that describes the multi-Weyl semimetals effectively. Setting $\hbar = 1$, the Hamiltonian reads 

\begin{equation}
	    H = C_s (k_y-sQ)\mathbb{I} + s\alpha\boldsymbol{\sigma}.\boldsymbol{n_k},
	    \label{eqn:low_eng_ham}
\end{equation}

where $\boldsymbol{n_k}$ = [$k_{\perp}^n \cos(n\phi),k_{\perp}^n \sin(n\phi),v_z(k_z-sQ)/\alpha$], $k_{\perp} = \sqrt{k_x ^2 + k_y ^2}$, $\boldsymbol{\sigma} = (\sigma_x,\sigma_y,\sigma_z)$ is the triad of Pauli matrices, $s=\pm 1$ characterizes the chirality of the Weyl point, $\mathbb{I}$ is the identity matrix, $C_s$ is the tilt parameter, and $\phi = \tan^{-1}{(k_y/k_x)}$. Here $\alpha$ generalizes the Fermi velocity in $k_x$-$k_y$ plane and is a dimensionless quantity (in our chosen units). The multi-Weyl nodes are separated by $2Q$ along the $k_z$ direction. $v_z$ acts as the Fermi velocity. Finally, $n$ is the monopole charge associated with the node. The dispersion relation for this model is given by 

\begin{equation}
	    E_{\pm}(\boldsymbol{k}) = C_s (k_y -sQ) \pm \sqrt{(k_z-sQ)^2 v_z^2 + \alpha^2(k_x^2+k_y^2)^n}.
	    \label{eqn:low_eng_ham_dispersion}
\end{equation}

Note the power dependence on $n$ with varying $k_x$ and $k_y$ -- near the Weyl point the energy bands follow a linear, quadratic and cubic behaviour, respectively, for $n=$ 1, 2 and 3. The parameter $C_s$ controls the tilt along $k_y$ direction. We have consistently used $s = -1$ and $Q = 0.5$ in this article. For these particular parameter values the bands will touch each other at $(0,0,-0.5)$ point in the momentum space. 

We begin our analysis by deriving analytical expressions for Berry curvature for general topological charge $n$. Using Equations \ref{eqn:berry_curvature}, \ref{eqn:low_eng_ham}, \ref{eqn:low_eng_ham_dispersion} we calculate the Berry curvature components along the three directions. These read

\begin{subequations}
    \begin{equation}
        \Omega_{\mp}^x =\pm \frac{n v_z k_x \alpha^2 (k_x^2 + k_y^2)^{n-1}}{2s[(k_z-sQ)^2 v_z^2 + \alpha^2(k_x^2+k_y^2)^n]^{\frac{3}{2}}},
        \label{eqn:omega_x}
    \end{equation}
    \begin{equation}
        \Omega_{\mp}^y =\mp \frac{n v_z k_y \alpha^2 (k_x^2 + k_y^2)^{n-1}}{2s[(k_z-sQ)^2 v_z^2 + \alpha^2(k_x^2+k_y^2)^n]^{\frac{3}{2}}},
        \label{eqn:omega_y}
     \end{equation}
         \begin{equation}
        \Omega_{\mp}^z = \pm \frac{\alpha^2 n^2 (k_z-sQ) s v_z (k_x^2 + k_y^2)^{n-1}}{2[(k_z-sQ)^2 v_z^2 + \alpha^2(k_x^2+k_y^2)^n]^{\frac{3}{2}}}.
        \label{eqn:omega_z}
    \end{equation}
    \label{eqn:bc_low_eng}
\end{subequations}

It is worth noting that, $\Omega_z$ has $n^2$ dependence compared to the $n$ dependence in $\Omega_x$ and $\Omega_y$. This plays a key role in determining relative magnitudes of different BCD components. Also note that these expressions reduce to the expected ones for $n=1$, i.e., the usual Weyl semimetal case.

The band dispersions along $k_y$ (for $k_x = k_z = 0$), with the value of the $z$ component of the Berry curvature superimposed on them for various tilt values are plotted in Fig.~\ref{fig:band_bcd}. We note that the Berry curvature is opposite for the valence and conduction bands. In case of no tilt ($C_s=0$), the inversion symmetry of the system is not broken and hence in a time-reversal symmetric system, such as ours, there is no Berry dipole. This can be attributed to the symmetric distribution of Berry curvature near the band edges. With the introduction of tilt along the $k_y$ direction, the inversion symmetry is broken and Berry curvature is now asymmetric at the band edges for all values of topological charge $n$. Hence peaks in the BCD appear, as shown in Fig.~\ref{fig:band_bcd}(d2)-(d3). We find that the maxima of the BCD is not exactly at the same energy where the Berry curvature is concentrated as the group velocity ($\frac{\partial E}{\partial k_y}$) vanishes at that point. So, an optimal energy where the product of group velocity and Berry curvature is maximum gives the maximum value of the dipole. As we will show, this optimal energy window can be investigated by analysis of the BCD density, as defined in Equation \ref{eqn:dipole_density}. Because of the absence of carriers in the band gap, the BCD is zero for energies within the gap. However, with larger tilts, a finite density of carriers is present invariably at all energy values for higher order band crossings and hence a non-zero BCD is found at all energies (except precisely at the band touching point). This can be seen clearly from Fig.\ref{fig:band_bcd}(d3). However, for the Weyl semimetal with topological charge $n=1$, there is always a zero BCD within the band gap. For higher order band crossings, Berry curvature is concentrated further away from the Brillouin zone centre (i.e., $k_y$ = 0) and the group velocity ($\frac{\partial E}{\partial k_y}$) is also higher near those points owing to the higher order dispersion. As a result, we discover that the BCD is highest for the case of $n = 3$, followed by $n= 2$ and $1$. With further increase in tilt, the bands have an even more asymmetric distribution of Berry curvature and larger group velocities, which lead to further increase in the magnitude of BCD. Other than $D_{yz}$, only the $D_{zy}$ component is non-zero amongst the off-diagonal ones in the BCD tensor. This component is smaller than the $D_{yz}$ component but follows a similar distribution where increasing tilt increases the magnitude of BCD.

\begin{figure}
	    \centering
	    \includegraphics[width=0.9\linewidth]{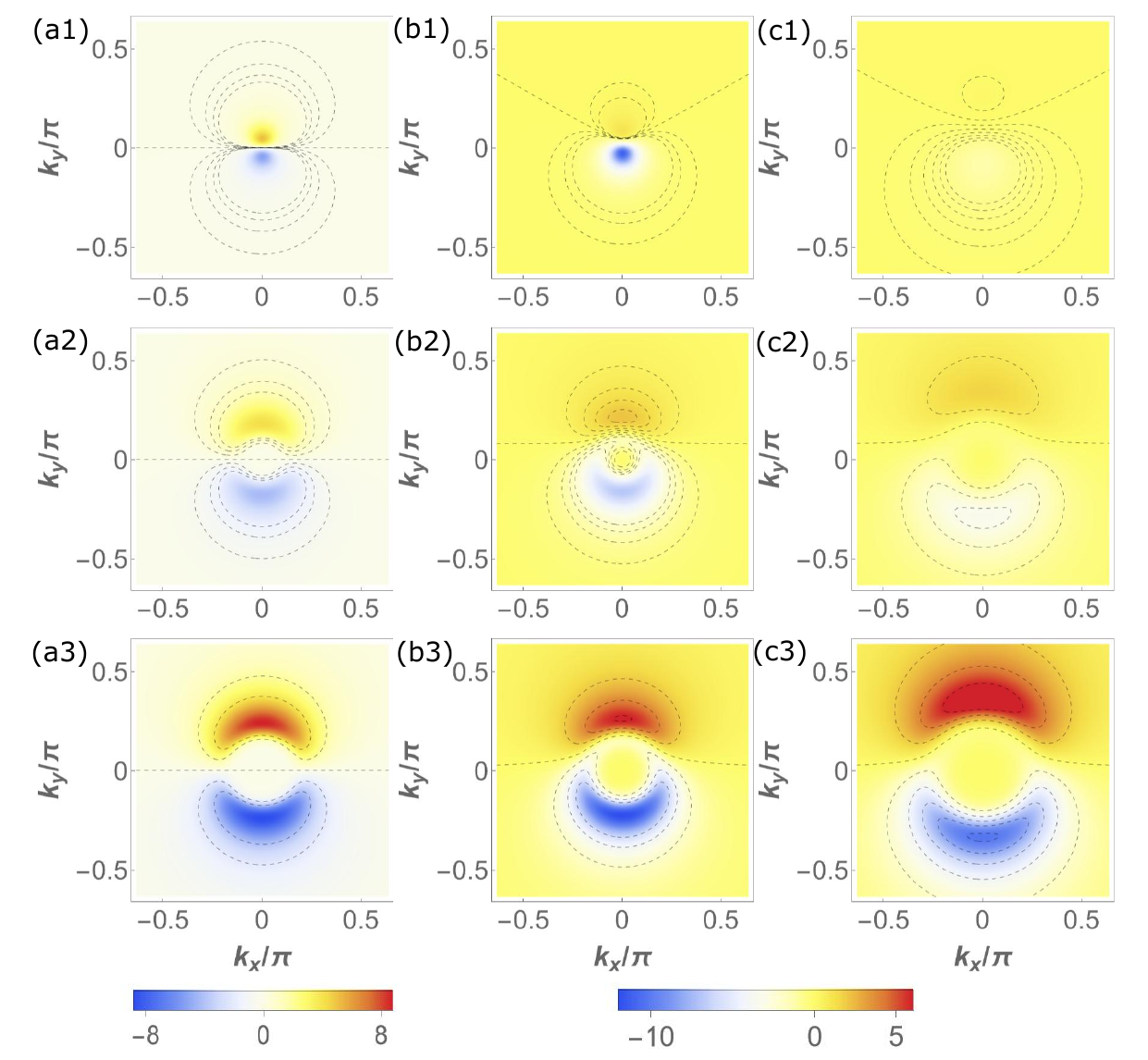}
	    \caption{\textbf{Berry curvature dipole densities with tilt.} Berry curvature dipole densities plotted in the $k_x-k_y$ plane (setting $k_z=0$) for (a1) $n=1$, $C_s=0$, (a2) $n=2$, $C_s=0$, (a3) $n=3$, $C_s=0$, (b1) $n=1$, $C_s=1$, (b2) $n=2$, $C_s=1$, (b3) $n=3$, $C_s=1$. For plots (c1)-(c3) we set $k_z=1$ for (c1) $n=1$, $C_s=1$, (c2) $n=2$, $C_s=1$, and (c3) $n=3$, $C_s=1$. Color scale for panels (b) and (c) are kept same for a better comparison. Dashed lines are contours for dipole density values close to zero. The distribution of dipole densities is anti-symmetric along $k_y$ for zero tilt and hence no net Berry curvature dipole is obtained. With the introduction of tilt this exact anti-symmetry disappears and as a result finite dipole originates. As we go further away from $k_z = 0$, the dipole density declines and spreads out suggesting a lower contribution to Berry curvature dipole from $k_x-k_y$ planes further away. Other parameters are $s=-1$, $Q=0.5$, $v_z=1$, $\alpha = 2$.}
	    \label{fig:dipole_density}
\end{figure}

To gain further insights into the nature of the BCD in multi-Weyl semimetals, we next turn to the BCD density. In Fig.~\ref{fig:dipole_density}, the dipole density, $d_{yz}$, as defined by Equation \ref{eqn:dipole_density} are plotted for all three values of $n$. When tilt is zero the distribution of dipole density is symmetric in a $k_x-k_y$ plane for all values of $k_z$ and $n$, hence cancels out to give a net zero contribution to BCD, as we present in Fig.~\ref{fig:dipole_density}(a). As we introduce a finite tilt, the dipole density distribution becomes asymmetric and gives a contribution from all $k_x-k_y$ planes which is seen in Fig.~\ref{fig:dipole_density}(b). We also note that the positive and negative values of $d_{yz}$ are also different, thereby resulting in a finite net value. We also find that the extremum points in the dipole density distribution are close to the multi-Weyl nodes. On the other hand, the points in the momentum plane which are away from multi-Weyl points contribute little to the BCD. For $n=3$, dipole density is highest in magnitude and has a larger momentum space spread as well. As a result, $n=3$ multi-Weyl semimetal produces the highest BCD. In the case of $n=1$, maximum value of dipole density is higher but it spans a very small momentum region, when compared to $n=2$ and $n=3$. So, the total BCD value for $n =2$ surpasses that of $n=1$, which produces the BCD variation that we show in Fig.~\ref{fig:band_bcd}. The BCD densities in Fig.\ref{fig:dipole_density}(c) are plotted for the same tilt as Fig.~\ref{fig:dipole_density}(b), but for $k_z = 1$ which is located further away from the location of the multi-Weyl points. We observe that the dipole density reduces. This decrease in dipole density is specially pronounced for $n=1$ and $n=2$, as shown in Fig.~\ref{fig:dipole_density}(c).

It is worth noting here that the dipole density, $d_{yz}$, is an odd function of $k_z$, while $d_{zy}$ is an odd function of $k_y$ for each of the bands for a three-dimensional material. So, the full three-dimensional integral will lead to net zero off-diagonal components of the BCD for this model, which is consistent with the recent findings of Ref.~\onlinecite{PhysRevB.103.245119}. However, as the authors have pointed out, this is not the case for many real materials and first-principles calculations show a high value of off-diagonal components of BCD in Weyl semimetal materials~\cite{zhang2018berry}. Thus, low energy models are insufficient to provide complete details of BCD in three-dimensional systems. However, as we have seen, they can help in understanding the general trends and their origin from two-dimensional subsystems. To get a complete understanding of BCD in multi-Weyl semimetals, we move on to tight-binding models in the next section, which can provide a more realistic description of real materials.

\subsection{Tight-binding model}

\begin{figure}
	    \centering
	    \includegraphics[width=1.0\linewidth]{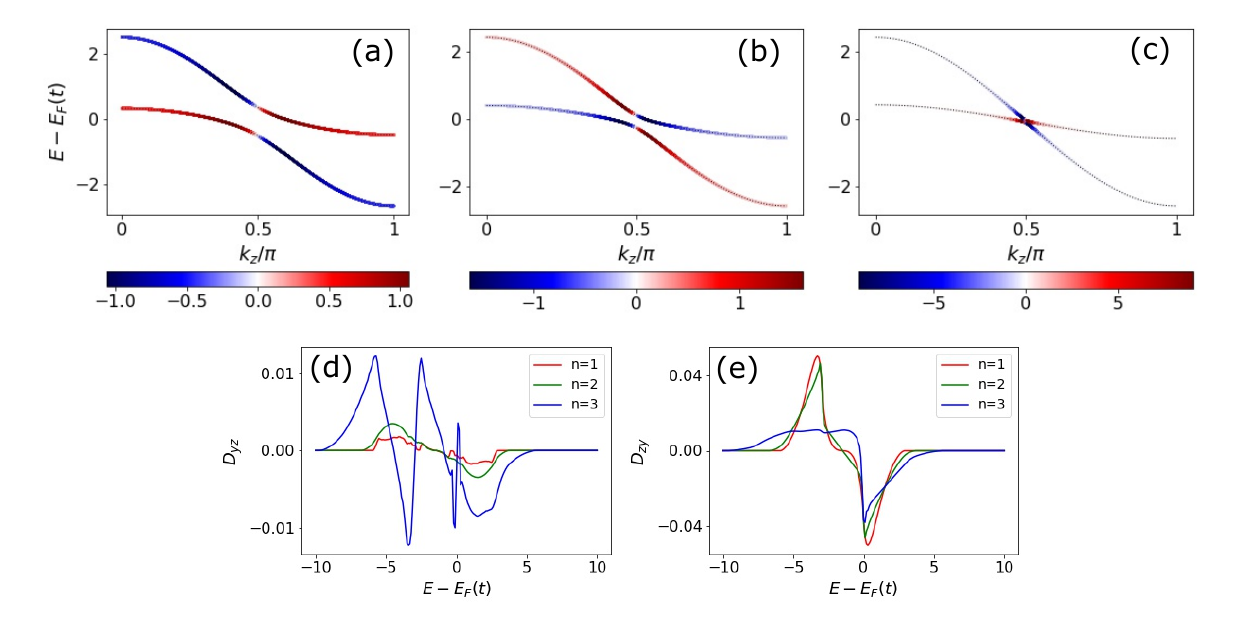}
	    \caption{\textbf{Band structure, Berry curvature and Berry curvature dipole for lattice models of multi-Weyl semimetals.} Band structure with the values of Berry curvature superimposed on it is plotted for the lattice Hamiltonian along $k_z$ direction with $k_y = 0.1 \pi$ and $k_x = 0.1 \pi$ for (a) $n=1$, (b) $n=2$, (c) $n=3$. Note the largest values of Berry curvature for highest topological charge $n=3$. Non-zero Berry curvature dipole components (d) $D_{yz}$  and (e) $D_{zy}$ are plotted for the same parameters. Relative peak values of Berry curvature dipole for $n = 1, 2$, $3$ match our predictions from the low-energy model for the $D_{yz}$ component. However, for the $D_{zy}$ component we observe a different trend because, for this case, the dominant factor for determining magnitude of Berry dipole is the $y$ component of the Berry curvature, which follows a different dependence on the topological charge. Other parameters are chosen to be $t_C=1.5$, $t = 1$ and $t_z =1$.}
	    \label{fig:tb_bcd}
\end{figure}

Motivated by the intriguing behavior of BCD in low energy models of multi-Weyl semimetals, we next study a lattice model for these systems. The tight-binding Hamiltonian reads~\cite{PhysRevB.104.075129}

\begin{equation}
    H^n = d_0^n + \boldsymbol{d}^n.\boldsymbol{\sigma},
    \label{eqn:tb_hamiltonian}
\end{equation}

where $\boldsymbol{\sigma}$ = ($\sigma_x,\sigma_y,\sigma_z$) are the Pauli matrices, $d_0^n$ and $\boldsymbol{d}^n$ are lattice periodic functions and $n$ is the topological charge of the multi-Weyl semimetal. The following parameters control the behaviour of the Hamiltonian: $t_C$ represents the tilt of the Weyl spectrum, while $t$ and $t_z$ are the hopping strengths. The lattice constant is taken to be unity. The components of $D^n = [d_0^n,\boldsymbol{d}^n]$ for $n=1,2,3$ are the following

\begin{subequations}
    \begin{equation}
        D^1 = [t_C(\cos{k_z} + \cos{k_x}-1) , t\sin{k_x},t \sin{k_y}, t_z \cos{k_z}],
        \label{eqn:tb_model_n1}
    \end{equation}
    \begin{equation}
        D^2 = [t_C(\cos{k_z} + \cos{k_x}-1), t (\cos{k_x}-\cos{k_y}), 2 t \sin{k_x}\sin{k_y},t_z \cos{k_z} ],
    \label{eqn:tb_model_n2}
    \end{equation}
\begin{equation}
    D^3 = [t_C(\cos{k_z} + \cos{k_x}-1), t \sin{k_x}(1-\cos{k_x}-3(1-\cos{k_y})), -t\sin{k_y}(1-\cos{k_y}-3(1-\cos{k_x})),t_z \cos{k_z} 
        ].
        \label{eqn:tb_model_n3}
\end{equation}    
\end{subequations}

In all the three cases the Weyl points are found at ($0,0,\pm\pi/2$). The magnitude of the tilt ($t_C$) determines whether the system is a type-I or a type-II Weyl semimetal. The dispersion relations for the different topological charge $n$ are as follows

\begin{subequations}
    \begin{equation}
        E^{n=1}_{\pm} = t_C (\cos{k_x}+\cos{k_z}-1) \pm [(t^2(2-\cos{2k_x}-\cos{2k_y})+ t_z^2(1+\cos{2k_z}))/2]^{1/2},
        \label{eqn:tb_dispersion_n1}
    \end{equation}
    \begin{equation}
    \begin{aligned}
        E^{n=2}_{\pm} = & t_C(\cos{k_x}+\cos{k_z}-1) \pm [(t^2(4-\cos{2k_x}-\cos{2k_y}+\cos{(2(k_x+k_y))}+\cos{(2(k_x-k_y))} \\
        &-2(\cos{k_x-k_y}+\cos(k_x+k_y)))+t_z^2(\cos{2k_z}+1))/2]^{1/2},
    \end{aligned}    
        \label{eqn:tb_dispersion_n2}
    \end{equation}
    \begin{equation}
    \begin{aligned}
        E^{n=3}_{\pm} = & t_C(\cos{k_x}+\cos{k_z}-1) \pm [t^2((\sin{k_x}(1-\cos{k_x}-3(1-\cos{k_y})))^2 \\
        & +(-t\sin{k_y}(1-\cos{k_y}-3(1-\cos{k_x})))^2)
         +(t_z \cos{k_z})^2]^{1/2}.
    \end{aligned}
        \label{tb_dispersion_n3}
    \end{equation}
\end{subequations}

For our numerical computations using the above tight-binding models, we employ the PythTB~\cite{pythtb} code to generate the lattice system. Then we use the Wannier-Berri package~\cite{Tsirkin2021,Destraz2020} for calculating the band structures, densities of states, Berry curvature and BCD. For the calculation of BCD a $100\times 100\times 100$ $k$-grid was used and convergence was checked.

We begin our investigation by examining the band structures for different multi-Weyl systems, as shown in Fig.~\ref{fig:tb_bcd}(a)-(c) along $k_z$ with $k_x = k_y = 0.1\pi$. The superimposed values of Berry curvature confirm that the Berry curvature is indeed concentrated near the Weyl points. The $z$ component of the Berry curvature is highest in magnitude for $n=3$ by some margin, which is the case for our low energy model as well. Further, the Berry curvature is opposite in sign for valence and conduction bands and also changes sign at the point at which the bands nearly touch.

The Hamiltonian described above has certain symmetry features that are important to understand which components of BCD will be non-zero. First of all it is time-reversal symmetric. Note that in this case anti-Hermitian complex conjugation is the time reversal operator. Inversion symmetry is broken for this system. It has mirror symmetries along $k_y$ and $k_z$. In presence of these mirror symmetries~\cite{PhysRevLett.115.216806}, only two components of BCD are expected to be non-zero, $D_{yz}$ and $D_{zy}$, and this is confirmed by our numerical calculations.

\begin{figure}
	    \centering
	    \includegraphics[width=1.0\linewidth]{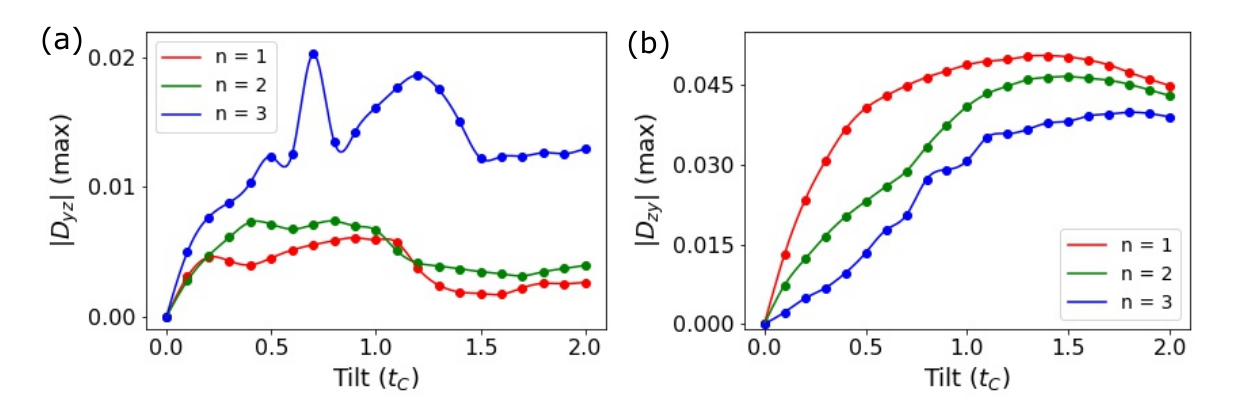}
	    \caption{\textbf{Maximum of Berry curvature dipole with tilt parameter.} Variation of maximum magnitude of Berry curvature dipole component (a) $D_{yz}$ and (b) $D_{zy}$ with tilt ($t_C$) for the lattice models of multi-Weyl semimetals for $n=1, 2$ and $3$. The maximum of Berry curvature dipole variation with tilt is different for the two components. However, both components start to saturate after a certain tilt value. This indicates that several factors come in play in determining the magnitude of Berry curvature dipole in multi-Weyl semimetals. Other fixed parameters are $t =1$ and $t_z = 1$.}
	    \label{fig:tilt_vary}
\end{figure}

The two non-zero components for all three topological charges are plotted in Fig.\ref{fig:tb_bcd}(d)-(e). The two components have an opposite relative magnitude of peaks of Berry dipole for the three topological charges. For $D_{yz}$, highest magnitude is observed for $n=3$, whereas magnitude is highest for $n=1$ for $D_{zy}$ component. These differences occur because of the difference in origin of the BCD in these two cases. For $D_{yz}$ the involved Berry curvature component is $\Omega_z$, which is perpendicular to the $k_x-k_y$ planes, in which multi-Weyl dispersion properties are present in our model. In other words, $\Omega_z$ is directly sensitive to the multi-Weyl nature of the dispersion as well as the higher topological charge. Besides $\Omega_z$, the group velocity plays a major role in determining the magnitude of Berry dipole here. It is different for the three topological charges and largest for $n=3$, again owing to the different dispersion. On the other hand, for $D_{zy}$, the relevant Berry curvature component is $\Omega_y$, which decreases with increase in topological charge. Since the $\sigma_x$ and $\sigma_y$ coefficients of Hamiltonian change for the three topological charges, behaviour of $\Omega_y$ is different than $\Omega_z$. This plays the major role in determining the BCD magnitude, as the relevant group velocity ($v_z$) is the same for all three values of $n$. Thus we uncover an unexpected variation in BCD sign and magnitude with topological charge, depending on the tensor component. In Fig.~\ref{fig:tilt_vary}(a), we plot the variation of the maximum of $D_{yz}$ with tilt and observe that for topological charge $n=3$, the dipole is greatest in magnitude for all tilt values. Topological charges $n=1$ and $n=2$ are close in value to each other. On the other hand, for $D_{zy}$ component, shown in Fig.~\ref{fig:tilt_vary}(b), we find that $n=1$ has the highest maximum BCD for all tilt values followed by $n=2$ and $n=3$. Our striking prediction of this anisotropy in the BCD should be directly verifiable in non-linear Hall measurements, as we will discuss later.

\section{Third order Hall response}

\begin{figure}
	    \centering
	    \includegraphics[width=0.95\linewidth]{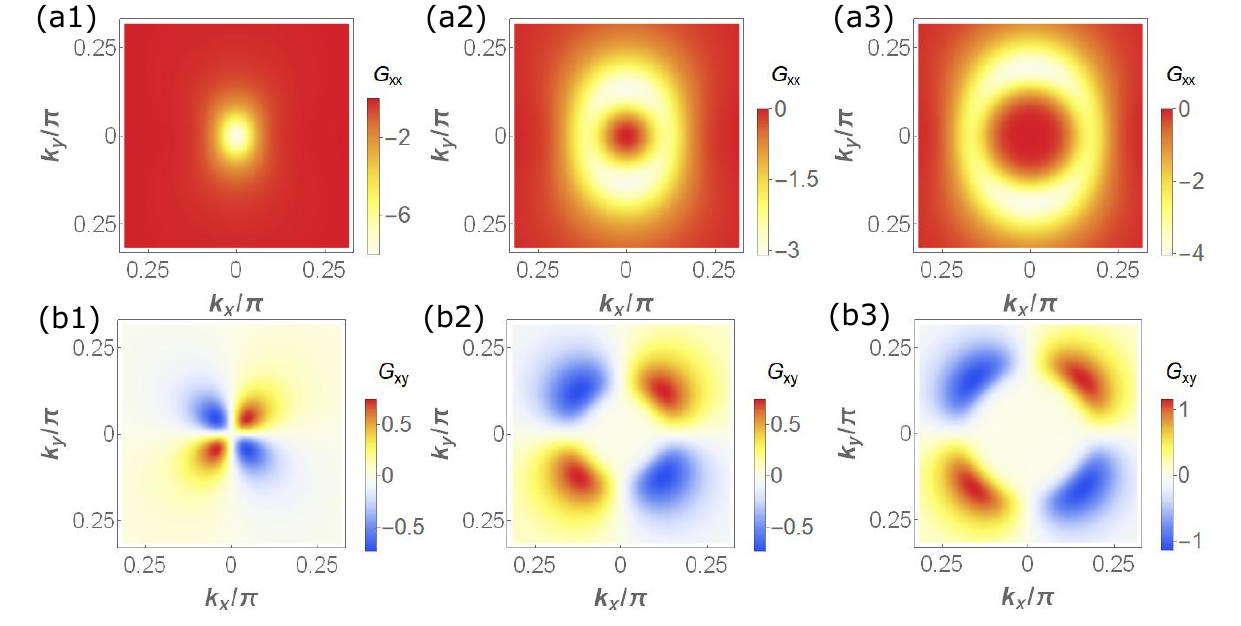}
	    \caption{\textbf{Berry connection polarizability tensor components.} Distribution of Berry curvature polarizability tensor components in the $k_x-k_y$ plane. Here $G_{xx}$ is plotted for (a1) $n=1$, (a2) $n=2$, (a3) $n=3$, and $G_{xy}$ is shown for (b1) $n=1$, (b3) $n=2$, (b3) $n=3$ for $s=-1$, $Q=0.5$ and $k_z = 0$. These are independent of the tilt of the system. Here $G_{xx}$ behaves like a monopole, while $G_{xy}$ exhibits a quadrupole-like pattern. Magnitudes of both components are peaked at positions where the Berry curvature peaks, i.e., where the band gap is minimal.}
	    \label{fig:BPT_comparison_2}
\end{figure}

After studying the second order non-linear Hall response in multi-Weyl semimetals, a natural question arises: can these systems show a higher order Hall response? To answer this question, we next investigate the third order Hall response in these systems. As we discussed before, the third order Hall response can be understood in the framework of the BCP tensor~\cite{liu2021berry,Lai2021}. We begin by deriving analytical expressions for the BCP components, for a general $n$, using our low energy Hamiltonian (see Equation \ref{eqn:low_eng_ham}). Using Equation \ref{eqn:bc_polarizability}, we find that the components of BCP are as follows

\begin{subequations}
    \begin{equation}
        G_{xx} = -\frac{n^2 \alpha^2 (k_x^2 + k_y^2)^{n-2} [k_x^2 k_z^2 v_z^2 + k_y^2(k_z^2 v_z^2+(k_x^2+k_y^2)^n \alpha^2)]}{4 [k_z^2 v_z^2+(k_x^2+k_y^2)^n \alpha^2]^{5/2}},
    \end{equation}
    \begin{equation}
        G_{yy} = -\frac{n^2 \alpha^2 (k_x^2 + k_y^2)^{n-2} [k_y^2 k_z^2 v_z^2 + k_x^2(k_z^2 v_z^2+(k_x^2+k_y^2)^n \alpha^2)]}{4 [k_z^2 v_z^2+(k_x^2+k_y^2)^n \alpha^2]^{5/2}},
    \end{equation}
    \begin{equation}
        G_{zz} = -\frac{v_z^2 \alpha^2 (k_x^2+k_y^2)^n}{4 [k_z^2 v_z^2+(k_x^2+k_y^2)^n \alpha^2]^{5/2}},
    \end{equation}
    \begin{equation}
        G_{xy} = G_{yx} = \frac{n^2 k_x k_y \alpha^4 (k_x^2 + k_y^2)^{2(n-1)})}{4 [k_z^2 v_z^2+(k_x^2+k_y^2)^n \alpha^2]^{5/2}},
    \end{equation}
    \begin{equation}
        G_{xz} = G_{zx} = \frac{k_x k_z n v_z^2 \alpha^2 (k_x^2 + k_y^2)^{(n-1)}}{4 [k_z^2 v_z^2+(k_x^2+k_y^2)^n \alpha^2]^{5/2}},
    \end{equation}
    \begin{equation}
        G_{yz} = G_{zy} = \frac{k_y k_z n v_z^2 \alpha^2 (k_x^2 + k_y^2)^{(n-1)}}{4 [k_z^2 v_z^2+(k_x^2+k_y^2)^n \alpha^2]^{5/2}}.
    \end{equation}
\end{subequations}
	
Note that all nine components of the BCP tensor are, in general, non-zero for multi-Weyl semimetals and depend on the topological charge $n$. Further, the off-diagonal components are symmetric, i.e., $G_{xy}=G_{yx}$, $G_{xz}=G_{zx}$, and $G_{yz}=G_{zy}$. In Fig.~\ref{fig:BPT_comparison_2}, we plot the different representative components of the BCP tensor in the $k_x-k_y$ plane. We notice several interesting features of the BCP tensor. The diagonal components $G_{xx}$ and $G_{yy}$ behave like a monopole for topological charge $n =1$ and are peaked at $k_x=k_y=0$, i.e., where the band gap is minimal. For topological charges $n = 2$ and $3$, these components still behave like a monopole, but are not peaked at the origin, but rather at a position slightly away from it. These peaks are in the same momentum region where the Berry curvature shows peaks for the corresponding topological charge, i.e., where the band gaps are minimal. Strikingly, the off-diagonal component, $G_{xy}$, has a quadrupole-like structure and shows a similar peak behaviour, where the peaks move further from $k_x=k_y=0$ with increase in the topological charge.	

\begin{figure}
	    \centering
	    \includegraphics[width=0.98\linewidth]{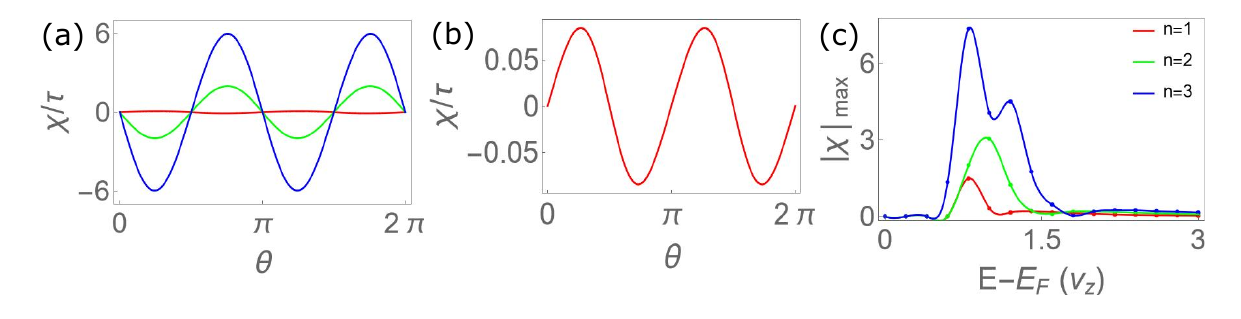}
	    \caption{\textbf{Third order conductivity tensor for multi-Weyl semimetals.} Angular dependence of the third order conductivity tensor. The different curves are for $n=1$ (red), $n=2$ (green) and $n=3$ (blue). (b) Zoomed-in plot of the curve for $n=1$ in panel (a). Third order conductivity increases substantially with increasing monopole charge. (c) The variation of the maximum conductivity with energy. It is near zero at very low and high Fermi energies and peaks at a moderate Fermi energy similar to the Berry curvature dipole distribution. Other parameters are chosen to be $C_s=0.5$, $v_z = 1$, $\alpha = 2$, $s =-1$, $Q=0.5$, $E-E_F = 1.1$, and $k_z=0$.}
	    \label{fig:third_cond}
\end{figure}

Calculating the full conductivity tensor analytically for a three-dimensional material is a formidable task with many components of $G$ and $\chi$ contributing to it and leading to cumbersome expressions. Rather, to gain more insight, we calculate the conductivity tensor for the special case considering $k_z = 0$. For simplifying the symmetry discussions, we assume $Q = 0$. In this case, the multi-Weyl points are located on the mirror line $M_x$ = $\sigma_x$, such that $M_x H(k_x,k_y) M_x^{-1} = H(-k_x,k_y)$. When the applied electric field is along or perpendicular to this mirror line, there will not be any third order Hall current generated. This means that $\chi_{yyyx}$, $\chi_{xxxy}$ and other such permutations with three $y$ and one $x$ components or three $x$ and one $y$ components will be zero and will not contribute to the conductivity tensor. On the other hand, an in-plane electric field, defined as $\boldsymbol{E} = E (\cos \theta, \sin \theta)$ will produce a third order response $j_H^{(3)} = \boldsymbol{j}^{(3)}\cdot\hat{\boldsymbol{n}}$, where $\hat{\boldsymbol{n}} = (-\sin\theta, \cos \theta)$ is the normal to $\boldsymbol{E}$. The third order contribution to the Hall conductivity is then determined by the conductivity tensor $\chi_{H} = \frac{j_H^{(3)}}{E^3}$. For our system, with the prescribed symmetries we find the third order conductivity tensor as

\begin{equation}
    \chi_H(\theta) = (-\chi_{11} + 3 \chi_{21}) \sin^3 \theta \cos \theta + (\chi_{22} - 3\chi_{12}) \sin \theta \cos^3 \theta ,
    \label{eqn:model_cond_3}
\end{equation}

where $\theta$ is measured from the mirror line, and the shorthand notations stand for $\chi_{11} = \chi_{xxxx}$, $\chi_{22} = \chi_{yyyy}$, $\chi_{12} = (\chi_{xxyy} + \chi_{xyxy} + \chi_{xyyx})/3$ and $\chi_{21} = (\chi_{yyxx} + \chi_{yxyx} + \chi_{yxxy})/3$. We have evaluated the third order conductivity tensor numerically for different multi-Weyl semimetals and the results are shown in Fig.~\ref{fig:third_cond}. For the special case of no tilt, $\chi_H$ is zero for all values of $\theta$ due to symmetry reasons discussed above. For a finite tilt, we observe a non-zero value of the conductivity tensor and the sign for $n=1$ is opposite to the cases of  $n=2$ and $n=3$. Remarkably, we find that the magnitude of $\chi$ is highest for $n = 3$ followed by $n=2$ and $n=1$. This is due to higher group velocity, less symmetry and sharper features in $G_{ab}$ for higher topological charges. For all values of topological charge, $\chi_H$ varies with the direction of applied field (i.e., with $\theta$) with a periodicity of $\pi$. It vanishes when $\theta$ is a multiple of $\pi/2$, which is consistent with our symmetry analysis. In Fig.~\ref{fig:third_cond}(c), we have presented the maximum value of the conductivity tensor with the Fermi energy. It turns out to show a variation similar to the BCD, where there is a peak at moderate Fermi energy but becomes nearly zero at higher and lower values of Fermi energy. Further, multi-Weyl semimetals with $n=3$ have the highest third order response followed by $n=2 $ and $n=1$. Therefore, our calculations show that multi-Weyl semimetals with higher topological charges can be suitable platforms for observing enhanced higher order Hall response.

\section{Experimental considerations}

Here, we estimate the magnitude of second order non-linear Hall response that can be measured in experiments. First, we need to calculate the conductivity tensors in Equation \ref{eqn:2nd_order_conductivity}. In typical experimental setups~\cite{Ma2019,kang2019nonlinear}, the relaxation time, $\tau$, is of the order of picoseconds and the ac frequency can be varied between 10-1000 Hz. So, the frequency dependence in denominator can be neglected as $\omega\tau \ll 1$. Using these values, the non zero components of the conductivity tensor are

\begin{subequations}
\begin{equation}
        \chi_{xyy} = -\chi_{yyx} = 0.1846 D_{yz},
        \label{chi_Dyz}
\end{equation}
\begin{equation}
    \chi_{xzz} = -\chi_{zzx} = -0.1846 D_{zy}.
\end{equation}
\end{subequations}

Using the previously mentioned relation, $J_a^{2\omega} = \chi_{abc} E_b E_c$, we can find the second order Hall current. For simplicity, we consider three cases where the current is confined in a cartesian plane, namely $xy$, $yz$ or $zx$. This gives us the following expressions for the current

\begin{subequations}
\begin{equation}
    J_y^{2\omega} = \chi_{yyx} E_y E_x \quad \textrm{when}\quad \boldsymbol{E} = (E_x, E_y, 0),
    \label{eqn:exp_j_y}
\end{equation}
\begin{equation}
    J_x^{2\omega} = \chi_{xyy} E_y^2 + \chi_{xzz} E_z^2 \quad \textrm{when}\quad \boldsymbol{E} = (0, E_y, E_z),
    \label{eqn:exp_j_x}
\end{equation}
\begin{equation}
    J_z^{2\omega} = \chi_{zzx} E_z E_x \quad \textrm{when}\quad \boldsymbol{E} = (E_x, 0, E_z).
    \label{eqn:exp_j_z}
\end{equation}
\end{subequations}

Consider a typical sample dimension to be $10\times 6\times 2$ $\mu$m$^3$ and the resistivity to be isotropic in all directions such that $\rho_x = \rho_y = \rho_z = 50$ $\mu \Omega$cm. Let us consider $V_x= 1$ V and $V_y = 1$ V. This choice gives us the current density, $J_y^{2\omega} =-0.1846 D_{yz} \frac{V_y V_x}{L_x L_y} = -3.077\times 10^9 D_{yz}$ A/m$^2$. This, in turn, produces a voltage, $V_y^{2\omega} = -1.8462 \times 10^4 \rho_y D_{yz}$ V. Therefore, we obtain the ratio to be $\frac{V_y^{2\omega}}{(V_x^{\omega})^2} = -9.231 \times 10^{-3} D_{yz}$ V$^{-1}$. This ratio is $16.3 \times 10^{-6}$ V$^{-1}$ for $n=1$, $31.9 \times 10^{-6}$ V$^{-1}$ for $n=2$, and $112.8 \times 10^{-6}$ V$^{-1}$ for $n=3$. These are well within the reach of current experimental techniques and can clearly distinguish our proposal of topological charge dependence of the non-linear Hall signal. It can also be noted that, for the typical parameters we used, the current $J_x^{2\omega}$ (Equation \ref{eqn:exp_j_x}), will have a overall higher magnitude of response. The above discussion also makes it clear that we can orient the applied electric fields to measure second harmonics that are dependent only on one of the BCD components or both the components.

\section{Summary and outlook}

In summary, we studied Berry curvature multipole physics in multi-Weyl semimetals. Using a low-energy model, we calculated the general expressions for Berry curvature and its dependence on topological charge, $n$. We discovered the occurrence of Berry curvature dipole in multi-Weyl semimetals. Our low-energy model predicted a general trend of increasing BCD magnitude with topological charge. With inspiration from low-energy model results, we used a tight-binding lattice model to study the dipole in a more realistic setting. We found two different variations in the its components with topological charges. While one component ($D_{yz}$) increases with topological charge the other one ($D_{zy}$) decreases with it. From studying their origins, we concluded that the magnitude of the Berry curvature dipole depends on size of the involved Berry curvature component (which, in turn, is dependent on the topological charge), the extent to which it is distributed in the momentum space and the group velocity of the relevant bands. Further, we analytically obtained the Berry connection polarizability tensor -- responsible for the third harmonic generation of the Hall signal -- for a low energy model and found all its components to be non-zero in multi-Weyl semimetals. We calculated the third order conductivity and showed that its magnitude increases with the underlying topological charge. We hope that our work can motivate exploration of Berry curvature multipole physics in multi-Weyl semimetals and help in characterizing new classes of multi-Weyl semimetal materials with higher topological charges.  

\section*{Acknowledgments}

We acknowledge useful discussions with N. B. Joseph, D. Varghese, A Bandyopadhyay, H. Liu, S. A. Yang,  S. Bhowal and N. A. Spaldin. S. R. thanks the Kishore Vaigyanik Protsahan Yojana (KVPY) for a fellowship. A. N. acknowledges support from the startup grant of the Indian Institute of Science (SG/MHRD-19-0001) and DST-SERB (project number SRG/2020/000153).


%

\end{document}